\documentclass[
aps,
prc,
reprint,
superscriptaddress,
amsmath,amssymb,
nofootinbib,
floatfix
]{revtex4-2}

\usepackage{graphicx}
\usepackage{bm}
\usepackage{booktabs}

\begin{document}

\title{Carbon-Reaction Constraints on Stellar Evolution:
White-Dwarf Heating, Carbon-Burning Products, and Black-Hole Masses}

\author{A.~M.~Mukhamedzhanov}
\affiliation{Cyclotron Institute, Texas A\&M University,
College Station, Texas 77843, USA}

\date{\today}

\begin{abstract}
The effects of ${}^{12}\mathrm{C}+{}^{12}\mathrm{C}$ fusion and
${}^{12}\mathrm{C}(\alpha,\gamma){}^{16}\mathrm{O}$ capture on
stellar heating, composition, and black-hole masses are examined.
Rates from a published Bayesian analysis of carbon fusion are
used in calculations with Modules for Experiments in Stellar
Astrophysics (MESA). In an accreting carbon--oxygen white dwarf,
the Bayesian rates move the point at which nuclear heating equals
combined neutrino and surface photon losses to slightly larger
stellar masses and higher central temperatures and densities.
At an accretion rate of $10^{-7}\,M_\odot\,\mathrm{yr}^{-1}$,
the Bayesian median gives a central temperature approximately
$2.57\%$ higher and a central density $18.9\%$ higher than the
calculation using the standard CF88 rate. This equality does not
by itself establish thermonuclear runaway. A separate calculation
follows carbon burning in an initially $20\,M_\odot$, $Z=0.0142$
star. Relative to the CF88 rate with MESA's standard division
between proton and alpha channels, the Bayesian-median rate
with the proton-to-alpha ratio of Yang et al. increases the
neon-20 mass remaining at central carbon depletion by $10.8\%$
and decreases the sodium-23 mass by $28.7\%$. These differences
combine changes in the total rate and branching; their numerical
uncertainties remain to be quantified. For helium capture on
carbon, the nuclear constraint
$S_{\rm tot}(300\,\mathrm{keV})=118\pm17\,\mathrm{keV\,b}$
is applied to the published $Z=10^{-5}$ stellar models of
Farag et al. Linear interpolation gives a maximum black-hole
mass below the pair-instability gap of approximately
$61.5$--$77.4\,M_\odot$ within the assumptions of those models.
New MESA calculations at $Z=0.00142$ give smaller bound masses
before collapse. Differences in the stellar assumptions prevent
attributing the difference between the two curves to metallicity alone.

\end{abstract}

\maketitle

\section{Introduction}
\label{sec:introduction}

Carbon links several stages in the lives of stars. It is produced
when three helium nuclei combine. A carbon nucleus can then capture
a helium nucleus to form oxygen, or two carbon nuclei can fuse
to produce heavier elements during a later burning stage.
The rates of these reactions affect the energy released inside
a star, its changing composition, and the mass that remains
at the end of its evolution. The role of nuclear burning in
massive-star evolution is reviewed by Woosley, Heger, and
Weaver~\cite{Woosley2002}. The stellar effects of
${}^{12}\mathrm{C}+{}^{12}\mathrm{C}$ have been investigated in
many studies of carbon burning, element production, and the
structure of massive stars, including those of Bennett et
al.~\cite{Bennett2012}, Pignatari et al.~\cite{Pignatari2013},
Chieffi et al.~\cite{Chieffi2021}, Monpribat et
al.~\cite{Monpribat2022}, and Gesu\`e et al.~\cite{Gesue}.
A recent review by Chieffi et al.~\cite{Chieffi2025} discusses
the nuclear uncertainties and their stellar consequences.
Complementary information on resonances relevant to carbon
fusion has been obtained by Nan, Wang, and collaborators
using thick-target inverse kinematics~\cite{NanWang2025}.
The importance of ${}^{12}\mathrm{C}(\alpha,\gamma){}^{16}\mathrm{O}$
for stellar evolution is reviewed by deBoer et al.~\cite{deBoer2017}.
Its effect on the maximum black-hole mass below the
pair-instability gap has been studied by Farmer et
al.~\cite{Farmer2020}, Woosley and Heger~\cite{WoosleyHeger2021},
and Farag et al.~\cite{Farag2022}.
Wang et al.~\cite{Wang2026} report evidence for low-spin
black holes with masses of $50$--$70\,M_\odot$ and discuss
the implications for the pair-instability mass cutoff.

Building on these studies, the present work evaluates the
stellar consequences of two recent nuclear inputs: the Bayesian
carbon-fusion rate of Ref.~\cite{MukCC2026} and the
${}^{12}\mathrm{C}(\alpha,\gamma){}^{16}\mathrm{O}$ constraint
of Ref.~\cite{MukBHM2026}. For carbon fusion, MESA calculations
quantify changes in the balance between nuclear heating and
combined neutrino and surface photon losses in an accreting
white dwarf. A separate
massive-star comparison examines the isotope masses remaining
after central carbon depletion, using the Bayesian median
together with the proton-to-alpha rate ratio of Yang et
al.~\cite{Yang}. For helium capture on carbon, the nuclear
constraint is applied to published stellar calculations at
$Z=10^{-5}$ and compared with new MESA calculations at
$Z=0.00142$ to examine the maximum black-hole mass below
the pair-instability gap. The purpose is to assess how these particular nuclear
inputs change the stellar predictions, while identifying the
limitations of each comparison.

Although both reactions involve carbon, they act at different
stages. The ${}^{12}\mathrm{C}(\alpha,\gamma){}^{16}\mathrm{O}$
reaction occurs during helium burning, when a carbon nucleus
captures an $\alpha$ particle (a helium nucleus) to form
oxygen-16, releasing energy as a photon. Its competition with the production of carbon from
three helium nuclei helps determine the amounts of carbon and
oxygen left when helium is exhausted. Carbon fusion acts later,
consuming this remaining carbon. Its principal channels produce
neon and a helium nucleus, or sodium and a proton. Further
reactions can then produce magnesium and other elements.
The carbon--helium reaction therefore helps establish the
starting composition for carbon burning, whereas carbon fusion
determines how that carbon is consumed.

Carbon fusion can supply heat in the dense interiors of
carbon--oxygen white dwarfs. These compact stars are supported
mainly by electron degeneracy pressure, which arises from
the quantum behavior of closely packed electrons. This pressure
depends less strongly on temperature than the pressure of an
ordinary gas. Compression, including that caused by the
accumulation of matter from a companion, can raise the density
and temperature until carbon fusion becomes important.
Nuclear reactions heat the star, while neutrinos and photons
carry energy away. Comparing these processes helps determine how
a change in the carbon-fusion rate affects the conditions
preceding ignition. Equality between heating and cooling
provides a useful reference point, but does not by itself
establish the onset of a thermonuclear runaway.%
\footnote{A thermonuclear runaway occurs when nuclear reactions
heat the matter faster than it can cool. The rising temperature
speeds up the reactions, causing still faster heating.
Depending on how the burning spreads and how much energy
is released, it can produce a flash, eject matter, or disrupt
the star. In a carbon--oxygen white dwarf, explosive carbon
burning can lead to a Type Ia supernova.}

\begin{figure*}[p]
\centering
\includegraphics[width=0.84\textwidth]{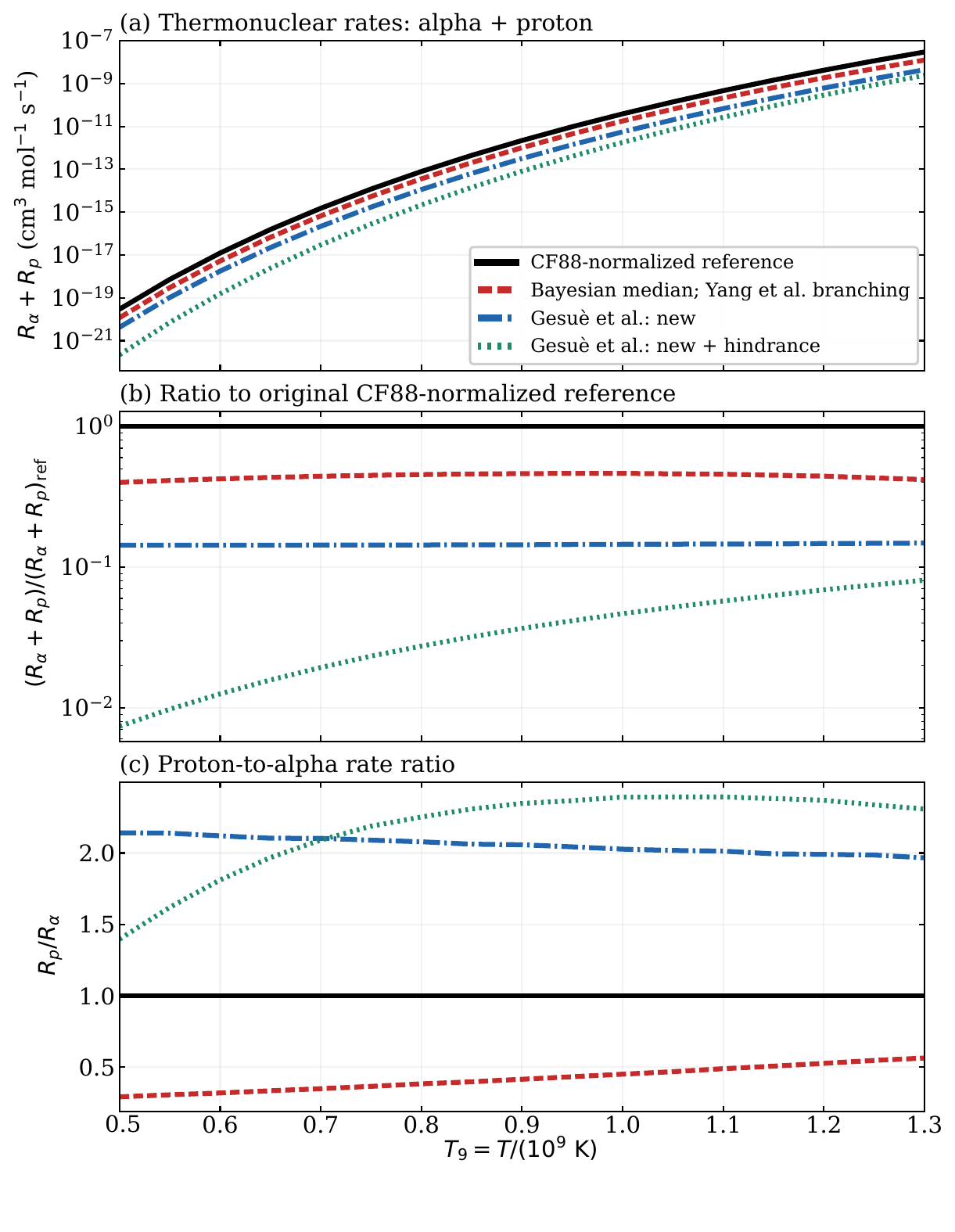}
\caption{Carbon-fusion rates and proton-to-alpha ratios.
The Bayesian-median rate is compared with the CF88 rate divided
between channels according to MESA and with the new and
new-plus-hindrance rates of Gesu\`e et al.~\cite{Gesue,GesueData}.
Upper panel: the sum $R_\alpha+R_p$. Middle panel: this sum
relative to the calculation using CF88. Lower panel: $R_p/R_\alpha$.
The Bayesian calculation uses the ratio of Yang et al.~\cite{Yang}
and is shown only through $T_9=1.3$. Its thermal integral covers
$1.0\leq E\leq3.5$ MeV; omitted contributions require assessment
near the upper plotted temperature. MESA gives $R_p/R_\alpha=1$
in this interval, differing from the original CF88 proportions.
Neutron and direct radiative channels are excluded from the sums.
Uncertainty bands are not shown.}
\label{fig:carbon_rates}
\label{fig:gesue_rate_comparison}
\end{figure*}

A recent Bayesian analysis of low-energy carbon-fusion
measurements provided an updated reaction rate and its
uncertainty range~\cite{MukCC2026}. This statistical approach
combines the available measurements to estimate the rates
consistent with the data and the assumptions of the analysis.
Figure~\ref{fig:carbon_rates} compares the resulting median
rate with a reference using the CF88 total rate~\cite{CF88}
and MESA branching, and with the
rates of Gesu\`e et al.~\cite{Gesue}; it also displays the
proton-to-alpha ratios, including that of Yang et al.~\cite{Yang}.
Here, $T_9=T/(10^9\,\mathrm{K})$. The Bayesian rate is plotted
only up to $T_9=1.3$.

\clearpage

The carbon-fusion part of this paper addresses two questions
with MESA stellar-evolution calculations. First, how does the
updated rate change the temperature and density at which
nuclear heating equals combined neutrino and surface photon
losses in an accreting
carbon--oxygen white dwarf? Calculations with the median,
lower, and upper rates distinguish the effect of replacing
CF88 from the uncertainty within the updated rate.
Second, how does replacing the CF88-total-rate reference with
the median rate change
the composition of a massive star during carbon burning?
For this comparison, two calculations begin from the same
model after helium burning and follow the depletion of central
carbon. The amounts and distributions of neon, sodium, and
magnesium are then compared. The reference uses MESA branching
with the CF88 total rate, whereas
the Bayesian-median sequence adopts the temperature-dependent
proton-to-alpha rate ratio of Yang et al.~\cite{Yang}.
The comparison therefore includes changes in both the total
fusion rate and its distribution between the charged-particle channels. These calculations describe
material remaining inside the star, rather than the amounts
eventually released into space.

Related studies have examined how revised carbon-fusion
rates affect the formation of carbon--oxygen, oxygen--neon,
and mixed-composition cores~\cite{Gesue}. The present
work connects the nuclear-rate comparison with both
white-dwarf heating and changes in the element abundances
during massive-star carbon burning.

The carbon--helium part of the paper concerns a different
consequence of the composition left after helium burning:
the mass available to form a black hole. In sufficiently
massive stellar cores, radiation can create electron--positron
pairs, reducing the pressure support and causing contraction.
The resulting rapid nuclear burning can drive pulses that
eject matter or can disrupt the star completely. The carbon
remaining after helium burning influences this later evolution.
The ${}^{12}\mathrm{C}(\alpha,\gamma){}^{16}\mathrm{O}$ rate
can therefore affect the maximum black-hole mass below the
range where complete stellar disruption prevents black-hole
formation. The black holes considered here form directly
from stars, rather than through previous black-hole mergers.

The constraint on $S_{\rm tot}(300\,\mathrm{keV})$ for the
${}^{12}\mathrm{C}(\alpha,\gamma){}^{16}\mathrm{O}$ reaction,
obtained in Ref.~\cite{MukBHM2026}, is applied to the
published stellar models of Farag et al.~\cite{Farag2022}
at $Z=10^{-5}$.  The results are also compared with MESA calculations
performed in the present work at $Z=0.00142$.  

Metallicity $Z$ is the mass
fraction of elements heavier than helium. It affects
stellar winds and the structure of the star, and hence
the mass remaining for black-hole formation. Because
the two sets of calculations also differ in other
assumptions, their differences cannot be attributed
to metallicity alone.

Together, these comparisons examine how the two
carbon-related reactions influence successive stages
of stellar evolution: the carbon and oxygen left
by helium burning, the heating and element production
during carbon burning, and the mass available for
black-hole formation.
\section{Bayesian $^{12}$C+$^{12}$C thermonuclear rate}
\label{sec:rates}

The adopted thermonuclear rates are tabulated in Ref.~\cite{MukCC2026}.
At each tabulated temperature, the inferred rate distribution is represented by the
16th, 50th, and 84th percentiles,
\begin{equation}
R_{16}(T_9),\qquad R_{50}(T_9),\qquad R_{84}(T_9).
\end{equation}
The median is $R_{50}$; at each temperature, the interval
from $R_{16}$ to $R_{84}$ contains the middle $68\%$ of
the inferred rate distribution. Here, $T_9=T/(10^9\,\mathrm{K})$ and
\begin{equation}
R(T_9)\equiv N_A\langle\sigma v\rangle
\end{equation}
is expressed in ${\rm cm^3\,mol^{-1}\,s^{-1}}$. Here, $N_A$ is
Avogadro's constant and $\langle\sigma v\rangle$ is the reaction
cross section times relative speed, averaged over the thermal
velocity distribution.  The CF88 rate,
$R_{\rm CF88}(T_9)$, is used as a reference \cite{CF88}.

The Bayesian rate was obtained by
integrating the nuclear input over
\begin{equation}
1.0\leq E\leq3.5~{\rm MeV}.
\end{equation}
For $^{12}\mathrm{C}+{}^{12}\mathrm{C}$, the approximate Gamow peak and
width are \cite{MukCC2026}
\begin{align}
E_0(T_9)&\simeq2.42\,T_9^{2/3}~{\rm MeV},\\
\Delta E(T_9)&\simeq1.05\,T_9^{5/6}~{\rm MeV}.
\end{align}
The Gamow window is the energy range contributing most strongly to
fusion at a given temperature. Table~\ref{tab:gamow_window} gives
representative peak energies and upper window edges,
$E_{\rm upper}\simeq E_0+\Delta E/2$.
At $T_9=1.3$, the approximate upper edge reaches 3.54 MeV,
slightly above the 3.5 MeV integration limit. The window
indicates where contributions outside the integration range
may matter; it does not quantify their size. 

\begin{table}[tb]
\caption{Approximate Gamow-peak energy $E_0$ and Gamow
window $[E_0-\Delta E/2,\,E_0+\Delta E/2]$ for
${}^{12}\mathrm{C}+{}^{12}\mathrm{C}$~\cite{MukCC2026}.
Here, $T_9=T/(10^9\,\mathrm{K})$; energies are given
in the center-of-mass system.}
\label{tab:gamow_window}
\begin{ruledtabular}
\begin{tabular}{ccc}
$T_9$ & $E_0$ (MeV) & Gamow window (MeV)\\
\hline
0.5 & 1.52 & 1.23--1.82\\
0.7 & 1.91 & 1.52--2.30\\
1.0 & 2.42 & 1.90--2.95\\
1.2 & 2.73 & 2.12--3.34\\
1.3 & 2.88 & 2.23--3.54\\
\end{tabular}
\end{ruledtabular}
\end{table}

\subsection{Direct comparison with the rates of Gesu\`e et al.}
\label{subsec:rate_comparison_gesue}

The two charged-particle channels provide a direct comparison
between the Bayesian input and the rates of Gesu\`e et
al.~\cite{Gesue}. Figure~\ref{fig:gesue_rate_comparison} uses
their published numerical tables~\cite{GesueData}, including
their new rates and their new rates with a model for additional
low-energy fusion suppression (hindrance). It covers
$0.5\leq T_9\leq1.3$ without extrapolating those tables.
The summed quantity is $R_\alpha+R_p$; neutron and direct
radiative channels are not included in this sum. The reference combines the CF88 total rate with the division between channels in MESA's \texttt{rate\_c12c12npa} routine.
For $T_9\leq3$, this routine assigns
$b_p=b_\alpha=(1-b_n)/2$, where $b_n$ is the neutron
branching fraction, and therefore gives $R_p/R_\alpha=1$.
The custom channel tables supplied to the reference calculation
retain these proportions. They differ from the original CF88
proportions, $R_\alpha:R_p=0.56:0.44$.

At $T_9=0.8$, the Bayesian-median alpha-plus-proton rate is
$3.584\times10^{-14}\,\mathrm{cm^3\,mol^{-1}\,s^{-1}}$,
compared with $1.133\times10^{-14}$ and
$2.166\times10^{-15}$ for the new and new-plus-hindrance
rates of Gesu\`e et al., respectively. The corresponding
proton-to-alpha ratios are approximately 0.384 for the
Yang et al.~\cite{Yang} branching used with the Bayesian
rate, and $2.079$ and $2.252$ for the two Gesu\`e et al.
prescriptions.  

At $T_9=1.3$, the summed rates $R_\alpha+R_p$ are
$1.250\times10^{-8}$, $4.420\times10^{-9}$, and
$2.408\times10^{-9}\,\mathrm{cm^3\,mol^{-1}\,s^{-1}}$
for the Bayesian median, the new rates of Gesu\`e et al.,
and their new-plus-hindrance rates, respectively.
Here, $R_\alpha$ denotes the rate for
${}^{12}\mathrm{C}+{}^{12}\mathrm{C}\rightarrow
{}^{20}\mathrm{Ne}+\alpha$, and $R_p$ denotes the rate for
${}^{12}\mathrm{C}+{}^{12}\mathrm{C}\rightarrow
{}^{23}\mathrm{Na}+p$.
The three prescriptions differ in both the summed rate
$R_\alpha+R_p$ and the proton-to-alpha rate ratio
$R_p/R_\alpha$.

The comparison at higher temperatures is also affected by
the energy range used to calculate each rate. The Bayesian
rate integral was evaluated over $1.0\leq E\leq3.5$ MeV,
whereas the prescription of Gesu\`e et al. includes
higher-energy cross sections. Table~\ref{tab:gamow_window}
shows that at $T_9=1.3$, the upper edge of the estimated
Gamow window reaches 3.54 MeV, slightly above the Bayesian
integration limit. The window alone does not determine
the size of the omitted contribution, which also depends
on the cross sections and any resonances above 3.5 MeV.
Differences between the rates near the upper plotted
temperature therefore cannot be attributed solely to
their low-energy extrapolations. Extending the Bayesian
rate to higher temperatures requires cross sections above
3.5 MeV and a new thermal integration.

\section{Nuclear energy-generation rate}
\label{sec:epsilon}

For matter of density $\rho$ and carbon mass fraction $X_C$, the molar abundance of $^{12}\mathrm{C}$ per unit mass is
\begin{equation}
Y_C=\frac{X_C}{A_C},\qquad A_C\simeq12\,\mathrm{g\,mol^{-1}}.
\end{equation}
The number of fusion events per unit volume and time is
\begin{equation}
r_{12,12}
=\frac{1}{2}\rho^2N_A
\left(\frac{X_C}{A_C}\right)^2R(T_9).
\end{equation}
If $Q_{\rm eff}$ is the energy deposited as heat per fusion
event, expressed in ergs, the heating rate per unit mass
from carbon fusion is
\begin{equation}
\boxed{
\epsilon_{\rm C+C}(\rho,T)=
\frac{\rho X_C^2}{2A_C^2}N_A R(T_9)Q_{\rm eff}
}.
\label{eq:epsilon}
\end{equation}
For calculations performed at the same temperature, density,
composition, $Q_{\rm eff}$, and screening enhancement,
\begin{equation}
\frac{\epsilon_i}{\epsilon_{\rm CF88}}
=
\frac{R_i}{R_{\rm CF88}},
\qquad i=16,50,84.
\label{eq:epsilon_ratio}
\end{equation}
Here, $R_i$ denotes the ${}^{12}\mathrm{C}+{}^{12}\mathrm{C}$
reaction rate from Ref.~\cite{MukCC2026}, with
$i=16,50,84$ corresponding to the 16th, 50th (median),
and 84th percentiles, respectively.

As a representative numerical example, at
\begin{equation}
\rho=10^7~{\rm g\,cm^{-3}},\qquad
T_9=0.80,\qquad X_C=0.50,
\end{equation}
the rates are
\begin{align}
R_{16}&=3.373\times10^{-14},&
R_{50}&=3.584\times10^{-14},\\
R_{84}&=3.823\times10^{-14},&
R_{\rm CF88}&=7.887\times10^{-14},
\end{align}
in ${\rm cm^3\,mol^{-1}\,s^{-1}}$.  Thus,
\begin{equation}
\frac{\epsilon_{50}}{\epsilon_{\rm CF88}}
=\frac{R_{50}}{R_{\rm CF88}}
=0.4544.
\end{equation}
The Bayesian median therefore produces about $45\%$ of the CF88
nuclear heating at the same temperature, density, and composition.
This illustrative ratio also assumes the same $Q_{\rm eff}$.
This expression illustrates the contribution from carbon fusion;
the full network heating also includes subsequent reactions.
Changing the proton-to-alpha branching can change the heat
deposited per fusion event and the subsequent reaction sequence.

\section{Effect of the carbon-fusion rate and branching on element production}
\label{sec:carbon_element_production}

The influence of ${}^{12}\mathrm{C}+{}^{12}\mathrm{C}$ fusion on
stellar composition was investigated with MESA r26.4.1 and the
75-isotope network \texttt{mesa\_75.net}. The reference
calculation uses the CF88 total rate~\cite{CF88}, divided
between the proton and alpha channels according to MESA's
\texttt{rate\_c12c12npa} routine. Its custom channel tables
assign equal proton and alpha rates for $T_9\leq3$.
The revised calculation uses the Bayesian-median rate of
Ref.~\cite{MukCC2026} and the temperature-dependent
proton-to-alpha ratio of Yang et al.~\cite{Yang}.
The Yang ratio is applied only to the revised calculation.
Numerical settings and endpoint properties are provided in
Appendix~\ref{app:carbon_numerics}.

Both sequences start from the same saved model after central
helium burning. The star initially had a mass of $20\,M_\odot$
and metallicity $Z=0.0142$. At the starting point, winds have reduced the
mass to approximately $15.0460\,M_\odot$, and the central
${}^{12}\mathrm{C}$ mass fraction is approximately $0.2485$.
Both calculations use the same mixing and mass-loss prescriptions.

\subsection{Implementation of the proton-to-alpha branching}

For the revised sequence, the ratio of thermonuclear rates is
\begin{equation}
q(T_9)=\frac{R_p(T_9)}{R_\alpha(T_9)}.
\label{eq:yang_branching}
\end{equation}
This is a ratio of thermally averaged rates, rather than
cross sections at a single collision energy. It was obtained
by digitizing the red curve in Fig.~8 of Yang et al.~\cite{Yang}.
The digitized values are approximately $0.293$, $0.452$, and
$0.527$ at $T_9=0.5$, $1.0$, and $1.2$, respectively.
Linear interpolation is used between tabulated points.
Below $T_9=0.5$, $q$ is held fixed at $0.29$; this continuation
is an explicit extrapolation. The adopted table is an
approximate reconstruction, not an author-supplied table.
Further implementation details are given in
Appendix~\ref{app:yang_implementation}.

Yang et al. calculated their thermonuclear channel-rate ratio using their energy-dependent branching prescription together with the CF88 assumption of an energy-independent total-fusion modified astrophysical factor, $S_{\rm tot}^{*}=3\times10^{16}\ \mathrm{MeV\,b}$. Applying their $q(T_9)$ to the present Bayesian total therefore
assumes that this ratio can also be used with that total.
The individual channel rates are not recalculated by integrating
the energy-dependent channel fractions with the Bayesian $S(E)$.
This procedure does not provide a Bayesian probability
distribution for each channel rate.

Let $R_{\alpha+p}^{(50)}$ denote the sum of the alpha and proton
rates in the original Bayesian-median input. The revised rates are
\begin{align}
R_\alpha^{(50,Y)}&=\frac{R_{\alpha+p}^{(50)}}{1+q},\\
R_p^{(50,Y)}&=\frac{qR_{\alpha+p}^{(50)}}{1+q}.
\label{eq:yang_repartition}
\end{align}
Here, the superscript $(50,Y)$ denotes the Bayesian-median rate (50th percentile) combined with the proton-to-alpha branching ratio of Yang et al.
The revised branching changes how carbon fusion is divided between the alpha and proton channels while keeping their combined rate unchanged at every tabulated temperature. 
No neutron channel was included in the adopted carbon-fusion input. For each modified channel, the reverse reaction rate is scaled by the same factor as the forward rate. Magnesium production is calculated through the reaction network; no separate adjustment is made to the direct fusion of two carbon nuclei into magnesium-24 with gamma-ray emission. 

Both sequences were evolved to central carbon depletion, defined by
$X_{\mathrm{C},c}<10^{-3}$. The isotope inventories are compared at
the saved endpoints; numerical settings and endpoint properties are
provided in Appendix~\ref{app:carbon_numerics}.

\subsection{Composition profiles and retained isotope masses}

\begin{figure*}[t]
\centering
\includegraphics[width=\textwidth]{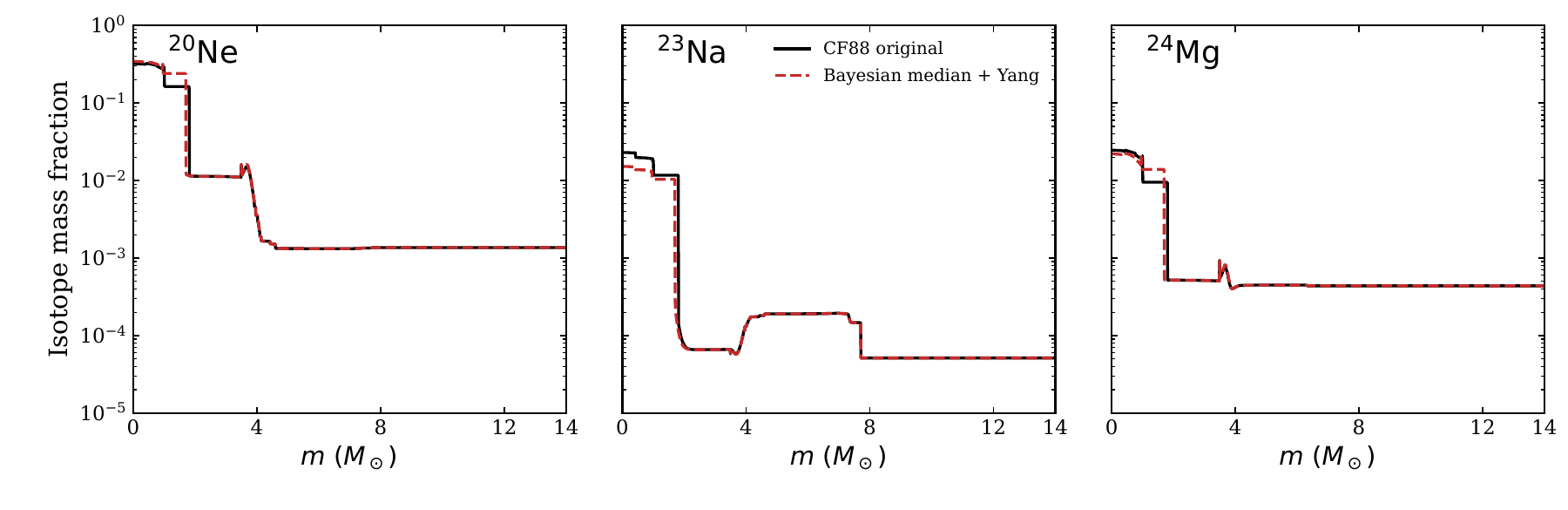}
\caption{Mass-fraction profiles of ${}^{20}\mathrm{Ne}$,
${}^{23}\mathrm{Na}$, and ${}^{24}\mathrm{Mg}$ at the saved
central-carbon-depletion endpoints. Black solid curves show
the CF88 total rate with MESA branching. Red dashed curves
show the Bayesian-median rate with the digitized Yang
proton-to-alpha branching~\cite{Yang}. Both sequences start
from the same post-helium-burning model of an initially
$20\,M_\odot$, $Z=0.0142$ star. The horizontal axes extend to
$m=14\,M_\odot$; the complete saved models extend to approximately
$14.77\,M_\odot$. The vertical axes are logarithmic. The final
central carbon fractions are $9.8657\times10^{-4}$ (CF88+MESA) and
$8.5030\times10^{-4}$ (Bayesian median with Yang branching).}
\label{fig:carbon_composition_profiles}
\end{figure*}

Figure~\ref{fig:carbon_composition_profiles} compares the
mass fractions of ${}^{20}\mathrm{Ne}$, ${}^{23}\mathrm{Na}$,
and ${}^{24}\mathrm{Mg}$ as functions of the enclosed stellar
mass $m$. The local mass fraction $X_i(m)$ is the fraction
of a small shell's mass contributed by isotope $i$.
The curves use the saved models at central carbon depletion.
Both calculations satisfy $X_{{}^{12}\mathrm{C},c}<10^{-3}$,
but their final central carbon fractions differ slightly.

The total mass of each isotope retained in the star is
\begin{equation}
M_i=\int_0^{M_\star}X_i(m)\,dm
\simeq\sum_k X_{i,k}\,\Delta m_k.
\label{eq:contained_isotope_mass}
\end{equation}
The integration uses every shell of each complete saved model,
including material outside the plotted interval. Here,
$\Delta m_k$ is the mass of shell $k$, $X_{i,k}$ is its isotope
mass fraction, and $dq_k=\Delta m_k/M_\star$ is its fraction
of the total stellar mass. The shell-mass sum
and the sum of the inventories over all 75 network species were
checked against the total stellar mass. Selected isotope masses
also agree with the integrated quantities in the profile headers.
These masses describe the material remaining in the star.
They do not give the change from the initial composition or
the amounts eventually ejected.

The percentage difference relative to the calculation using the CF88 rate
and MESA branching is
\begin{equation}
\Delta_i=100\%\,
\frac{M_i^{(50,Y)}-M_i^{\mathrm{ref}}}{M_i^{\mathrm{ref}}}.
\label{eq:isotope_relative_change}
\end{equation}
Here, $M_i^{\mathrm{ref}}$ is the isotope mass in the calculation
using the CF88 rate, and $M_i^{(50,Y)}$ is its mass in the
Bayesian-median calculation with the Yang ratio.
Table~\ref{tab:carbon_isotope_masses} gives selected isotope masses.

\begin{table}[t]
\caption{Retained isotope masses at the actual saved
carbon-depletion endpoints, for the calculation using the CF88 rate and MESA branching and Bayesian
median with Yang branching. The final column is defined in
Eq.~(\ref{eq:isotope_relative_change}). These masses include the whole star, rather than ejected
material. The effects of the different final carbon abundances
and the numerical resolution have not yet been quantified.
Percentages are calculated before rounding the isotope masses.}
\label{tab:carbon_isotope_masses}
\begin{ruledtabular}
\begin{tabular}{lrrr}
Isotope & CF88+MESA & Median + Yang & Change (\%)\\
 & $M_i/M_\odot$ & $M_i/M_\odot$ & \\
\hline
${}^{12}\mathrm{C}$ & 0.91172 & 0.89122 & $-2.25$\\
${}^{16}\mathrm{O}$ & 2.87581 & 2.85352 & $-0.78$\\
${}^{20}\mathrm{Ne}$ & 0.48088 & 0.53270 & $+10.78$\\
${}^{23}\mathrm{Na}$ & 0.03158 & 0.02251 & $-28.70$\\
${}^{24}\mathrm{Mg}$ & 0.03664 & 0.03624 & $-1.10$\\
${}^{25}\mathrm{Mg}$ & 0.01811 & 0.01872 & $+3.38$\\
${}^{26}\mathrm{Mg}$ & 0.02288 & 0.02427 & $+6.07$\\
${}^{27}\mathrm{Al}$ & 0.01189 & 0.01044 & $-12.21$\\
\end{tabular}
\end{ruledtabular}
\end{table}

\begin{figure}[t]
\centering
\includegraphics[width=\columnwidth]{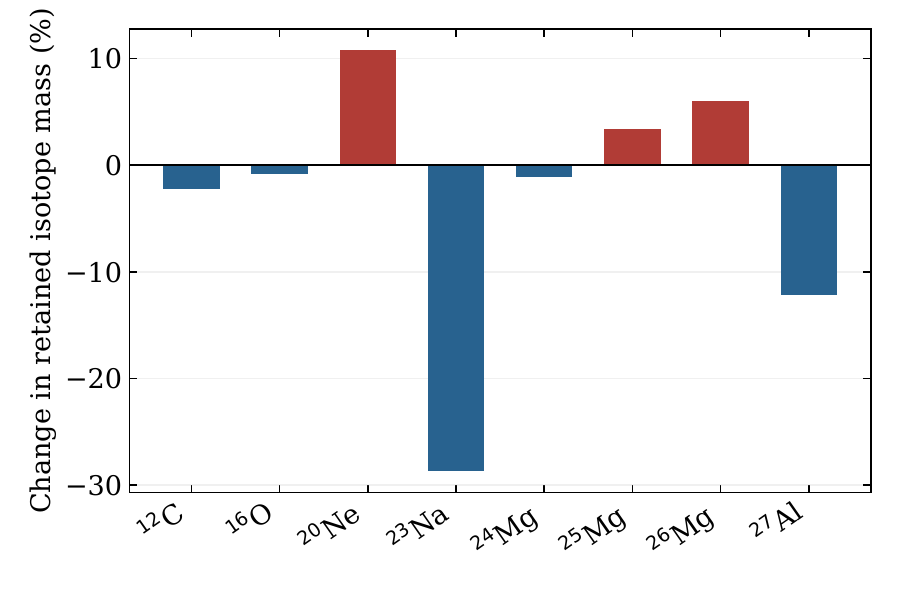}
\caption{Percentage change in retained isotope mass for Bayesian
median with Yang branching relative to the calculation using the CF88 rate and MESA branching, using
Eq.~(\ref{eq:isotope_relative_change}) and
Table~\ref{tab:carbon_isotope_masses}. Positive and negative bars
indicate increased and decreased inventories, respectively.
The bars compare two prescriptions for the total rate and
branching; they are not Bayesian uncertainty intervals or
ejected yields.}
\label{fig:retained_isotope_mass_change}
\end{figure}

The retained ${}^{20}\mathrm{Ne}$ mass increases by $10.78\%$,
whereas ${}^{23}\mathrm{Na}$ decreases by $28.70\%$.
The changes in ${}^{24}\mathrm{Mg}$, ${}^{25}\mathrm{Mg}$, and
${}^{26}\mathrm{Mg}$ are $-1.10\%$, $+3.38\%$, and $+6.07\%$,
respectively. Carbon-12, oxygen-16, and aluminum-27 decrease by
$2.25\%$, $0.78\%$, and $12.21\%$. 

The increased neon and decreased sodium are consistent with
the revised division of the rate between the alpha and proton
channels.
However, subsequent captures, mixing, and structural evolution
also contribute. Because both the total rate and the prescribed
branching differ between the sequences, the comparison does not
isolate the contribution of either change. Separating these effects would require additional calculations
that change only one of them at a time. In addition, the reference
uses MESA's proportions for the proton and alpha channels,
which differ from those in the original CF88 publication. The quoted abundance differences
therefore apply to the calculation using the CF88 rate and MESA branching; they do not establish
the changes relative to a calculation using the published CF88
branching. That comparison requires a separate reference run.

The quoted differences are preliminary. Both saved models have
central carbon fractions below the depletion threshold, but their
final fractions differ. Comparisons at the same central carbon
abundance and with finer numerical resolution are needed to
quantify these effects. The numerical uncertainty in the isotope
mass differences has therefore not been established. Large
percentage changes in isotopes present in negligible amounts
are not taken as evidence of significant production.

\subsection{Relation to the calculations of Gesu\`e et al.}

Gesu\`e et al.~\cite{Gesue} examined separately revised
proton- and alpha-channel carbon-fusion rates, with and without
a low-energy hindrance extrapolation, in stellar models
including those with an initial mass of $20\,M_\odot$. 
Table~2 of Gesu\`e et al.~\cite{Gesue} lists the temperature
and density at carbon ignition for three rate prescriptions:
CF88, the new rate, and the new rate with hindrance.
In this order, the ignition temperatures are
$\log_{10}(T/\mathrm{K})=8.87$, $8.90$, and $8.93$.
The corresponding densities are
$\log_{10}(\rho/[\mathrm{g\,cm^{-3}}])=5.24$, $5.45$,
and $5.62$. The carbon-burning lifetimes are 1203, 414, and 99 yr.
They also find changes in later shell locations and core compactness.

This provides relevant context for the sensitivity of stellar
structure to carbon-fusion inputs. However, their separately
constructed channel rates differ from the present prescription,
which divides the Bayesian alpha-plus-proton rate sum between
the two channels using the proton-to-alpha ratio of Yang et al. 
The present study compares isotope abundance profiles and the
total mass of each isotope remaining in the star at central
carbon depletion. Gesu\`e et al.~\cite{Gesue} instead report
conditions at carbon ignition and the structure of burning
shells at later stages. These results cannot be compared
directly because they describe different properties and stages
of stellar evolution. A direct quantitative comparison would
require both studies to use the same nuclear inputs and
stellar assumptions and to compare the same quantities at
the same evolutionary stage.

\section{White-dwarf MESA calculations}
\label{sec:wd}

The astrophysical consequences of the Bayesian
${}^{12}\mathrm{C}+{}^{12}\mathrm{C}$ rate were investigated
with MESA calculations of a carbon--oxygen white dwarf
accreting material from a companion white dwarf in a binary
system~\cite{Paxton2011,Jermyn2023}. Four evolutionary calculations of the accreting white dwarf
were performed:  
one using the CF88 carbon-fusion rate adopted as the standard
rate in MESA  and three using the 16th, 50th, and 84th percentiles
of the Bayesian rate distribution. The three Bayesian
calculations are denoted Bayes16, Bayes50, and Bayes84,
respectively. The
proton-to-alpha branching ratio of Yang et al. was not
applied to these sequences.

These calculations examine how the adopted carbon-fusion
rate affects the conditions at which nuclear heating
equals the combined energy losses through neutrino
emission and photon emission from the surface of the
accreting white dwarf. Neutrinos interact only weakly
with matter and escape from the stellar interior,
carrying energy away. Photons also remove energy
by escaping from the stellar surface. However, near the heating--cooling balance, the energy carried away by photons is only about $2.1$--$2.6\%$ of
that carried away by neutrinos. Neutrino emission therefore
dominates the energy losses in these calculations.
The principal comparison uses the same accretion rate
for all four sequences:
\begin{equation}
\dot M=10^{-7}\,M_\odot\,{\rm yr}^{-1}.
\end{equation}
The balance between nuclear heating and the combined
energy losses through neutrino and photon emission
is characterized by
\begin{equation}
L_{\rm nuc}=L_\nu+L_\gamma.
\label{eq:global_balance}
\end{equation}
Here, $L_{\rm nuc}$ is the energy generated by nuclear
reactions per unit time throughout the white dwarf,
$L_\nu$ is the energy carried away by neutrinos per unit
time, and $L_\gamma$ is the energy emitted as photons
from the stellar surface per unit time. This equality
provides a common comparison point for the four
calculations; it does not by itself establish thermal
equilibrium or the onset of runaway burning.

For each calculation, the difference
\begin{equation}
D_{\rm g}=L_{\rm nuc}-L_\nu-L_\gamma
\end{equation}
was evaluated as the star evolved. The balance point was
estimated by linear interpolation of $D_{\rm g}$ between
adjacent saved models on opposite sides of zero. The same
interpolation fraction was used for the stellar mass, age,
and logarithms of the central temperature and density.
The result depends on the spacing of the saved models;
numerical convergence has not been established.
Compression caused by accretion and heat transport between
layers also affect the star's thermal evolution. The
luminosity equality therefore does not establish whether
burning will run away in a particular region.

\begin{table*}[!t]
\caption{\label{tab:global_balance}
White-dwarf properties when nuclear heating equals the
combined neutrino and photon losses,
$L_{\rm nuc}=L_\nu+L_\gamma$, for the standard MESA
${}^{12}\mathrm{C}+{}^{12}\mathrm{C}$ rate and the Bayesian
16th-, 50th-, and 84th-percentile rates. All sequences use
$\dot M=10^{-7}\,M_\odot\,{\rm yr}^{-1}$.
Values are estimated by linear interpolation of
$L_{\rm nuc}-L_\nu-L_\gamma$ between adjacent saved models.
The quantity $t$ is the stellar age recorded in MESA.
The displayed digits do not imply established numerical
convergence.}
\begin{ruledtabular}
\begin{tabular}{lcccc}
Rate & $M/M_\odot$ & $\log_{10}(T_c/{\rm K})$ &
$\log_{10}(\rho_c/[{\rm g\,cm^{-3}}])$ & $t$ (yr)\\
\hline
Standard & 1.4118173 & 8.652903 & 10.704100 &
$8.7137138\times10^{7}$\\
Bayes16 & 1.4126659 & 8.664712 & 10.783994 &
$8.7145624\times10^{7}$\\
Bayes50 & 1.4126191 & 8.663913 & 10.779215 &
$8.7145155\times10^{7}$\\
Bayes84 & 1.4125659 & 8.663217 & 10.774034 &
$8.7144624\times10^{7}$\\
\end{tabular}
\end{ruledtabular}
\end{table*}

\begin{table*}[!t]
\caption{\label{tab:local_balance}
Central temperature, density, and energy quantities at
$\epsilon_{\rm nuc,c}=\epsilon_{\nu,c}$ for the Bayesian
$^{12}\mathrm{C}+{}^{12}\mathrm{C}$ rates.  The Bayesian results are given here; the standard-rate central
balance is listed in Table~\ref{tab:standard_mdot_balance}.}
\begin{ruledtabular}
\begin{tabular}{lcccc}
Rate &
$\log_{10}(T_c/{\rm K})$ &
$\log_{10}(\rho_c/[{\rm g\,cm^{-3}}])$ &
$C_P$ (${\rm erg\,g^{-1}\,K^{-1}}$) &
$\epsilon_{\rm nuc,c}=\epsilon_{\nu,c}$ (${\rm erg\,g^{-1}\,s^{-1}}$)\\
\hline
Bayes16 & 8.61007 & 10.75204 & $1.3863\times10^{7}$ & $1.709\times10^{3}$\\
Bayes50 & 8.60936 & 10.74719 & $1.3874\times10^{7}$ & $1.693\times10^{3}$\\
Bayes84 & 8.60860 & 10.74203 & $1.3886\times10^{7}$ & $1.676\times10^{3}$\\
\end{tabular}
\end{ruledtabular}
\end{table*}

\begin{table*}[!t]
\caption{\label{tab:standard_mdot_balance}
Central temperatures and densities at
$\epsilon_{\rm nuc,c}=\epsilon_{\nu,c}$ (loc) and
$L_{\rm nuc}=L_\nu+L_\gamma$ (glob), using the standard
CF88 rate in MESA. The central condition is interpolated
using the logarithm of the rate ratio; the whole-star
condition uses $L_{\rm nuc}-L_\nu-L_\gamma$.
Numerical convergence of these estimates has not been established.}
\begin{ruledtabular}
\begin{tabular}{ccccc}
$\dot M$ ($M_\odot\,{\rm yr}^{-1}$) &
$\log_{10}(T_{c,\rm loc}/\mathrm{K})$ &
$\log_{10}(T_{c,\rm glob}/\mathrm{K})$ &
$\log_{10}(\rho_{c,\rm loc}/[\mathrm{g\,cm^{-3}}])$ &
$\log_{10}(\rho_{c,\rm glob}/[\mathrm{g\,cm^{-3}}])$\\
\hline
$3\times10^{-8}$ & 8.57622 & 8.62777 & 11.32795 & 11.35081\\
$5\times10^{-8}$ & 8.58645 & 8.64015 & 11.02124 & 11.04741\\
$1\times10^{-7}$ & 8.59832 & 8.65290 & 10.67099 & 10.70410\\
\end{tabular}
\end{ruledtabular}
\end{table*}
Compared with the calculation using the standard MESA rate,
the differences in white-dwarf mass when total nuclear
heating equals combined neutrino and surface photon losses are
\begin{align}
\Delta M_{16}&=8.49\times10^{-4}\,M_\odot,\\
\Delta M_{50}&=8.02\times10^{-4}\,M_\odot,\\
\Delta M_{84}&=7.49\times10^{-4}\,M_\odot,
\end{align}
and the corresponding delays are approximately
\begin{align}
\Delta t_{16}&=8.49\times10^3~{\rm yr},\\
\Delta t_{50}&=8.02\times10^3~{\rm yr},\\
\Delta t_{84}&=7.49\times10^3~{\rm yr}.
\end{align}
These satisfy
\begin{equation}
\Delta M\simeq\dot M\,\Delta t.
\end{equation}

At this balance point, the calculation using the
Bayesian-median rate has a central temperature approximately
$2.57\%$ higher and a central density $18.9\%$ higher than
the calculation using the standard CF88 rate.
The quoted delays measure differences in the time of this
luminosity equality; they are not predicted explosion delays.

The white-dwarf mass at this heating--cooling balance is
largest for Bayes16, followed by Bayes50 and Bayes84, and
smallest for the standard MESA rate. Among the Bayesian
calculations, the lower fusion rate (Bayes16) requires the white dwarf
to accumulate more material and reach greater compression
before nuclear heating can match the combined neutrino
and surface photon losses.

\subsection{Heating and cooling at the stellar center}
\label{subsec:local_balance}

A separate comparison identifies the point at which nuclear
heating equals neutrino cooling at the center of the white dwarf:
\begin{equation}
\epsilon_{\rm nuc,c}=\epsilon_{\nu,c}.
\label{eq:local_balance}
\end{equation}
Here, $\epsilon_{\rm nuc,c}$ and $\epsilon_{\nu,c}$ are the
nuclear heating and neutrino cooling rates per unit mass
at the center.

Neutrinos produced in the central region escape from the
white dwarf and directly remove energy from that region.
Photons, however, are repeatedly absorbed and scattered
inside the star. They transfer energy between neighboring
layers rather than escaping directly from the center.
Electron conduction also transfers heat between layers.
The cooling caused by these transport processes depends
on the energy leaving a region minus the energy entering it.
The surface photon luminosity measures the energy escaping
from the entire star and does not specify the cooling
rate at its center.

Heat transport is not included in this particular comparison
of energy terms; this does not mean that it is omitted from
the stellar evolution calculation or that it is negligible. Equation~(\ref{eq:local_balance}) therefore
identifies equality between nuclear heating and neutrino
cooling only. It does not establish complete thermal
balance, which also depends on heat transport and
compressional heating.

The point of equality was estimated by interpolation
between adjacent saved models. Table~\ref{tab:local_balance}
uses linear interpolation of
$\epsilon_{\rm nuc,c}-\epsilon_{\nu,c}$, whereas
Table~\ref{tab:standard_mdot_balance} uses linear interpolation
of $\log_{10}(\epsilon_{\rm nuc,c}/\epsilon_{\nu,c})$.
The resulting estimates depend on the spacing of the saved
models and the interpolation method. Their numerical
convergence has not been established.

The central heat capacity per unit mass at constant pressure,
$C_P$, was calculated using MESA's equation of state, which
relates pressure, temperature, density, and composition.

The heat capacities at the three Bayesian balance points differ
by only about $0.2\%$. The changes in the balance temperature
and density mainly reflect the adopted reaction rates, with the
other assumptions held fixed.

A heating timescale based only on these two central terms is
\begin{equation}
t_{\rm heat}=
\frac{C_PT_c}{\epsilon_{\rm nuc,c}-\epsilon_{\nu,c}}.
\end{equation}
It tends to infinity as the denominator approaches zero
from above and is undefined at exact equality. This does
not imply that the complete local heating rate vanishes:
heat transport and compression also contribute to the
temperature evolution.

\subsection{Dependence on the accretion rate}
\label{subsec:mdot}

The rate at which the white dwarf gains material affects
how quickly it is compressed and heated. To examine how
this changes the conditions at which nuclear heating
equals cooling, calculations using the standard MESA
carbon-fusion rate were compared at three accretion rates:
\begin{equation}
\dot M=
3\times10^{-8},\quad
5\times10^{-8},\quad
10^{-7}\ M_\odot\,{\rm yr}^{-1}.
\end{equation}
These values specify the mass added to the white dwarf
each year, expressed in solar masses.

\clearpage

\section{The carbon--helium reaction and the maximum black-hole mass}
\label{sec:bhm}

The reaction ${}^{12}\mathrm{C}(\alpha,\gamma){}^{16}\mathrm{O}$
affects black-hole formation indirectly. A larger rate converts
more carbon into oxygen during helium burning, whereas a
smaller rate leaves more carbon for the next burning stage.
The resulting changes in composition affect the subsequent
temperature and density evolution of the core and its
susceptibility to instability caused by electron--positron
pair creation.

This section examines how the lower boundary of the
pair-instability black-hole mass gap---the maximum black-hole mass below the gap---depends on the total $S_{\rm tot}(300\,{\rm keV})$ factor for the 
${}^{12}\mathrm{C}(\alpha,\gamma){}^{16}\mathrm{O}$ reaction.

For this reaction, the adopted nuclear constraint is~\cite{MukBHM2026} 
\begin{equation}
S_{\rm tot}(300\,\mathrm{keV})=118\pm17\,\mathrm{keV\,b}.
\label{eq:S_conservative_intro}
\end{equation}
This interval is obtained from an $R$-matrix analysis
using the latest $\alpha$-removal asymptotic normalization
coefficients (ANCs) for the subthreshold $1^{-}$ and $2^{+}$
states of ${}^{16}\mathrm{O}$, together with a Bayesian
analysis of its ground-state ANC. 
The adopted interval includes
uncertainties beyond the narrower
$116$--$122\,\mathrm{keV\,b}$ range associated with the
bound-state information used in that analysis.

Two metallicities are considered using MESA stellar-evolution
calculations. At $Z=10^{-5}$, the nuclear interval in
Eq.~(\ref{eq:S_conservative_intro}) is applied to the published
calculations of Farag et al.~\cite{Farag2022}. At $Z=0.00142$,
new MESA calculations performed in the present work explore
how the maximum black-hole mass depends on the adopted
reaction rate at higher metallicity. The inferred masses
are theoretical estimates that depend on the assumptions
of the stellar models, rather than direct measurements.

\subsection{Relation to published stellar calculations}
\label{subsec:published_grid}

Mehta et al.~\cite{Mehta2022} calculated how the lower boundary
of the pair-instability black-hole mass gap depends on the
${}^{12}\mathrm{C}(\alpha,\gamma){}^{16}\mathrm{O}$ reaction
rate. In Ref.~\cite{MukBHM2026}, their results were expressed
as a dependence of the maximum black-hole mass below the gap
on $S_{\rm tot}(300\,\mathrm{keV})$, using the approximate conversion
\begin{equation}
S_{\rm tot}(300\,\mathrm{keV})\simeq(140+20\sigma)\,\mathrm{keV\,b},
\label{eq:Mehta_conversion}
\end{equation}
where $\sigma$ is a signed, dimensionless parameter describing
the variation of the reaction rate within its adopted
uncertainty range. Applying the nuclear constraint  (\ref{eq:S_conservative_intro})    
on $\,S_{\rm tot}(300\,\mathrm{keV})\,$ to this relation gave an estimated
maximum black-hole mass interval of about
$61$--$75\,M_\odot$.  Here, $\sigma$ labels the rate variation and does not denote
a reaction cross section.

The present low-metallicity estimate instead uses the
high-resolution, nonrotating helium-star calculations of Farag
et al.~\cite{Farag2022} at $Z=10^{-5}$. Their reaction-rate grid
covers $-3\leq\sigma\leq3$, corresponding to
$77\leq S_{\rm tot}(300\,\mathrm{keV})\leq203\,\mathrm{keV\,b}$.
The conversion for this grid is
\begin{equation}
S_{\rm tot}(300\,\mathrm{keV})=(140+21\sigma)\,\mathrm{keV\,b}.
\label{eq:Farag_conversion}
\end{equation}
The $21\,\mathrm{keV\,b}$ spacing in this equation is kept
distinct from the approximate $20\,\mathrm{keV\,b}$ conversion
in Eq.~(\ref{eq:Mehta_conversion}). The two expressions describe
different published rate prescriptions.

For $Z=10^{-5}$, the analysis uses the existing stellar
calculations of Farag et al.~\cite{Farag2022}; no new
stellar models are calculated at this metallicity.
The adopted interval of $S_{\rm tot}(300\,{\rm keV})$  
in Eq. (\ref{eq:S_conservative_intro})  is used to identify the corresponding 
range of maximum black-hole masses in their results.

The masses are taken from Table~2 of Ref.~\cite{Farag2022}
for $f_{3\alpha}=1$. This means that the triple-$\alpha$
reaction rate, which governs the formation of carbon
from three helium nuclei, is kept at the reference value
used by Farag et al.

The conservative nuclear-physics interval
$101\leq S_{\rm tot}(300\,{\rm keV})\leq135\,{\rm keV\,b}$
from Ref.~\cite{MukBHM2026} is converted to the Farag et al.
rate parameter  $\sigma$  using
\begin{equation}
\sigma=
\frac{S_{\rm tot}(300\,{\rm keV})/({\rm keV\,b})-140}{21}.
\end{equation}
The lower limit, central value, and upper limit give
\begin{align}
S_{\rm tot}(300\,{\rm keV})=101\,{\rm keV\,b}
&\quad\Longrightarrow\quad \sigma\simeq-1.8571,\\
S_{\rm tot}(300\,{\rm keV})=118\,{\rm keV\,b}
&\quad\Longrightarrow\quad \sigma\simeq-1.0476,\\
S_{\rm tot}(300\,{\rm keV})=135\,{\rm keV\,b}
&\quad\Longrightarrow\quad \sigma\simeq-0.2381.
\end{align}

The corresponding maximum black-hole masses are estimated
by linear interpolation between neighboring entries in
Table~\ref{tab:Farag_points}:
\begin{equation}
M_{\rm BH,max}(\sigma)=M_1+
\frac{\sigma-\sigma_1}{\sigma_2-\sigma_1}(M_2-M_1),
\end{equation}
where $M_1$ and $M_2$ are the tabulated masses at
$\sigma_1$ and $\sigma_2$, respectively.

The published masses used for interpolation are listed
in Table~\ref{tab:Farag_points}.

\begin{table}[t]
\caption{\label{tab:Farag_points}
Published maximum black-hole masses from the
$f_{3\alpha}=1$, $Z=10^{-5}$ sequence in Table~2 of
Farag et al.~\cite{Farag2022}. The rate parameter
$\sigma$ is related to $S_{\rm tot}(300\,{\rm keV})$
by Eq.~(\ref{eq:Farag_conversion}).}
\begin{ruledtabular}
\begin{tabular}{cc}
$\sigma$ & $M_{\rm BH,max}$ ($M_\odot$)\\
\hline
$-2.0$ & 79.55\\
$-1.5$ & 71.84\\
$-1.0$ & 64.41\\
$-0.5$ & 63.03\\
$0.0$  & 60.07\\
\end{tabular}
\end{ruledtabular}
\end{table}

For the lower nuclear limit,
$S_{\rm tot}(300\,{\rm keV})=101\,{\rm keV\,b}$,
interpolation between $\sigma=-2.0$ and $-1.5$ gives
$M_{\rm BH,max}\simeq77.35\,M_\odot$.
At the central value,
$S_{\rm tot}(300\,{\rm keV})=118\,{\rm keV\,b}$,
interpolation between $\sigma=-1.5$ and $-1.0\,$  gives
$\,M_{\rm BH,max}\simeq65.12\,M_\odot$.
For the upper nuclear limit,
$\,S_{\rm tot}(300\,{\rm keV})=135\,{\rm keV\,b}$,
interpolation between $\sigma=-0.5$ and $0$ gives
$\,M_{\rm BH,max}\simeq61.48\,M_\odot$.
The larger capture rate therefore corresponds to the
smaller maximum black-hole mass.

Table~\ref{tab:BHM_summary} summarizes these results.
Unrounded values of $\sigma$ were used in the
interpolation.

\begin{table}[t]
\caption{\label{tab:BHM_summary}
Maximum black-hole masses obtained by applying the
adopted nuclear interval to the published Farag et al.
grid at $Z=10^{-5}$.}
\begin{ruledtabular}
\begin{tabular}{cccc}
$S_{\rm tot}(300\,{\rm keV})$ (keV\,b) & 101 & 118 & 135\\
\hline
$\sigma$ & $-1.8571$ & $-1.0476$ & $-0.2381$\\
$M_{\rm BH,max}$ ($M_\odot$) & 77.35 & 65.12 & 61.48\\
\end{tabular}
\end{ruledtabular}
\end{table}

Thus, for the published $Z=10^{-5}$ stellar models \cite{Farag2022},
\begin{equation}
\boxed{
\begin{aligned}
101 {\rm keV\,b}  \leq
S_{\rm tot}(300\,{\rm keV})
&\leq135 {\rm keV\,b},\\
61.48\lesssim
\frac{M_{\rm BH,max}}{M_\odot}
&\lesssim77.35.
\end{aligned}}
\label{eq:BHM_final}
\end{equation}
The mass interval $61.48$--$77.35\,M_\odot$ was obtained
by varying the adopted nuclear input over
$101\leq S_{\rm tot}(300\,{\rm keV})\leq135\,{\rm keV\,b}$
while keeping the assumptions of the published stellar
models fixed. Uncertainties associated with stellar winds,
mixing, and other reaction rates are not included.
 The two decimal places result from the
interpolation and do not imply an accuracy of
$0.01\,M_\odot$.

\subsection{Why the reaction rate affects the black-hole mass}
\label{subsec:physical_origin}

The carbon remaining after helium burning affects where
carbon burning occurs and how much heat it produces.
The resulting changes in temperature and density also
affect energy loss through neutrino emission.
These processes change the temperature,
density, and composition inside the core before pair instability.

Radiation makes an important contribution to the thermal state
of a massive stellar core. Its entropy per unit mass is
\begin{equation}
s_{\rm rad}=\frac{4aT^3}{3\rho},
\label{eq:radiation_entropy}
\end{equation}
where $a$ is the radiation constant. Entropy describes how thermal
energy is distributed among the available states of matter and
radiation. At a fixed temperature, the radiation contribution
per unit mass is larger at lower density. The full stellar
entropy also includes ions, electrons, and, where relevant,
electron--positron pairs.

At sufficiently high temperatures, photons can create
electron--positron pairs, converting some of the radiation
energy into particle rest mass. As a result, pressure rises
less strongly as the core is compressed. If this weakening
of pressure support affects a sufficiently large part of
the core, gravity drives rapid contraction.
The core contracts and heats. Rapid oxygen burning can reverse
the contraction and drive an outward pulse. If the released
energy disrupts the entire star, no stellar remnant is left.
If the star survives, a pulse can expel part of its mass, and
later contraction can produce further pulses. This is the
pulsational pair-instability supernova (PPISN)
regime~\cite{Woosley2017}.

The nuclear reaction therefore affects the remnant through a
sequence of physical changes:
\begin{align}
&S_{\rm tot}(300)\longrightarrow X_{\rm C}^{\rm He,dep}
\longrightarrow\text{later core burning}\nonumber\\
&\longrightarrow\text{pair-instability pulses}
\longrightarrow M_{\rm bound}.
\label{eq:nuclear_bh_chain}
\end{align}
Here, $X_{\rm C}^{\rm He,dep}$ is the carbon mass fraction at the
end of helium burning,  and $M_{\rm bound}$ is the mass that
remains held by gravity. The surviving core may later undergo
collapse. An approximate mass accounting is
\begin{equation}
M_{\rm BH}\simeq M_{\rm initial}-M_{\rm wind}
-M_{\rm PPISN}-\Delta M_{\rm collapse},
\label{eq:bh_mass_budget}
\end{equation}
where $M_{\rm wind}$ and $M_{\rm PPISN}$ are the masses lost
through winds and pulses. The term $\Delta M_{\rm collapse}$
includes any further ejected mass and energy carried away during
collapse, expressed as an equivalent mass. Thus, the bound
stellar mass need not equal the gravitational mass of the
resulting black hole.

\subsection{Dependence on stellar metallicity}
\label{subsec:metallicity}

Metallicity\footnote{In astronomy, metallicity refers to the
abundance of elements heavier than helium. The quantity $Z$
is the fraction of stellar mass in these elements.}
affects how much matter a star loses before reaching pair
instability. In hot massive stars, the outward flow of light can
accelerate gas into a stellar wind. Iron-group ions absorb light
at many wavelengths because they have numerous allowed electronic
transitions. An absorbed photon supplies both excitation energy
and an outward momentum kick to the ion. Electric forces and
collisions between charged particles share that momentum with
the surrounding gas. A larger metal abundance generally increases
the force supplied by spectral lines and strengthens mass loss.

At $Z=0$, this metal-line contribution is absent, although other
mass-loss mechanisms can still operate. At $Z=10^{-5}$ it is
weak, so stars can retain much of their mass. Stronger winds at
$Z=0.00142$ can remove more material before collapse.
Metallicity also affects the star's internal structure
and the growth of its helium and carbon--oxygen cores.
Its effect on the maximum black-hole mass must therefore
be determined from stellar-evolution calculations.
It cannot be reliably estimated by simply shifting
the mass curve upward or downward.

For a specified set of stellar assumptions, the dependence can
be written as
\begin{equation}
M_{\rm BH,max}=M_{\rm BH,max}
\left[S_{\rm tot}(300\,\mathrm{keV}),Z\right].
\label{eq:bh_rate_metallicity}
\end{equation}
In this notation, the quoted $S$ factor identifies an adopted
reaction-rate prescription. A single $S$-factor value does not
specify the complete temperature dependence of the rate.

\subsection{MESA calculations at $Z=0.00142$}
\label{subsec:mesa_ppisn_method}

The higher-metallicity calculations were performed with MESA.
The initial helium and heavy-element mass fractions were
$Y=0.99858$ and $Z=0.00142$. For each reaction-rate factor,
a range of initial helium-star masses, $M_{\rm He}$, was
examined to estimate the largest mass remaining after the pulses.

The tabulated de Boer reference rate~\cite{deBoer2017} was
multiplied by a constant reaction-rate factor $f$. The adopted normalization is
\begin{equation}
S_{\rm tot}(300\,\mathrm{keV})=140f\,\mathrm{keV\,b}.
\end{equation}
The reaction-rate factor $f$ specifies the normalization
of the adopted ${}^{12}\mathrm{C}(\alpha,\gamma){}^{16}\mathrm{O}$
rate relative to a reference rate:
\begin{equation}
R(T)=f\,R_{\rm ref}(T).
\end{equation}
Here, $f=1$ gives the reference rate, while $f<1$ and $f>1$
give smaller and larger rates, respectively. The factor $f$ is independent of temperature.  This changes the overall magnitude of the rate while keeping the shape of its temperature dependence unchanged.
The range of $f$ covers the adopted nuclear interval in
Eq.~(\ref{eq:S_conservative_intro}). These calculations
vary the overall magnitude of the reference rate; they do
not reproduce differences in the temperature dependence
of independently determined rates.

MESA follows the motion of stellar layers during rapid
burning and pulses. The saved models were examined to
estimate how much material remained gravitationally bound.

A layer was identified as potentially escaping when its
outward radial velocity $v_r$ exceeded the estimated local
escape speed:
\begin{equation}
v_r(m)>v_{\rm esc}(m)
=\sqrt{\frac{2Gm}{r(m)}}.
\end{equation}
Here, $G$ is the gravitational constant, $m$ is the mass
enclosed within the layer, and $r(m)$ is its radius.
This expression uses the enclosed mass in a spherical
approximation. It neglects the contribution of overlying
layers to the gravitational potential and is therefore
an approximate escape criterion.

The mass coordinate of the innermost layer satisfying
this condition was designated $m_{\rm cut}$. The mass
remaining bound to the star was then estimated as
\begin{equation}
M_{\rm bound}\simeq m_{\rm cut},
\end{equation}
where $M_{\rm bound}$ is the estimated bound stellar mass.
This approximation treats the material inside the boundary
as bound and the material outside it as ejected.

The estimate is based on the velocities in a saved model.
It does not by itself account for subsequent pressure
forces, energy exchange between layers, or material that
may fall back onto the star. It therefore does not directly
determine the final black-hole mass. When no layer satisfied
the velocity criterion, the last calculated stellar mass
was retained as a provisional upper limit on the black-hole mass  from that model, assuming no later accretion or merger.
 
 The calculations examined initial helium-star masses
between $50$ and $85\,M_\odot$. For each reaction-rate
factor $f$, the adopted  black-hole mass estimate is the
largest among the selected completed calculations at
that value of $f$. A low-mass model
left a small bound core, while an $85\,M_\odot$ model was
classified as completely disrupted. These outcomes depend on
the adopted rate and are not universal mass boundaries.
The original mass uncertainties were assigned as
$\pm1.5\,M_\odot$ for the leftmost point and
$\pm1.0\,M_\odot$ for the others. They are not statistical
confidence intervals or a demonstrated bound on all model errors.

At $Z=0.00142$, the plotted masses are provisional upper
estimates of the black-hole masses that could form from
the selected stellar models. Most calculations were
completed through the pulsational mass-loss phase,
yielding the final bound stellar mass before collapse.
The subsequent collapse into a black hole was not followed,
so any additional mass loss during collapse remains
unquantified. Such losses depend on the stellar structure
and rotation. Calculations of related progenitors suggest
that the losses may be small, although ejection of a few
solar masses cannot be excluded~\cite{Rahman2022}.

To represent possible mass loss during collapse, an additional
downward allowance of $0.02M$ is adopted for every plotted
mass estimate $M$ at $Z=0.00142$. Treating this allowance
and the original uncertainty $\delta_1$ as independent,
the upper and lower error-bar lengths are
\begin{equation}
\delta_+=\delta_1,\qquad
\delta_-=\sqrt{\delta_1^2+(0.02M)^2}.
\label{eq:bh_mass_errors}
\end{equation}
The plotted interval extends from $M-\delta_-$ to
$M+\delta_+$; the central value $M$ is unchanged.
For example, at $M=50\,M_\odot$ with
$\delta_1=1\,M_\odot$, the upper and lower error bars
are $1\,M_\odot$ and approximately $1.41\,M_\odot$.
The $2\%$ contribution is an assumed allowance, not an
uncertainty calculated for the present models or established
by Ref.~\cite{Rahman2022}. It does not cover all possible
collapse losses.

\subsection{Comparison with the published low-metallicity models}
\label{subsec:metallicity_comparison}

\begin{figure}[t]
\centering
\includegraphics[width=\columnwidth]{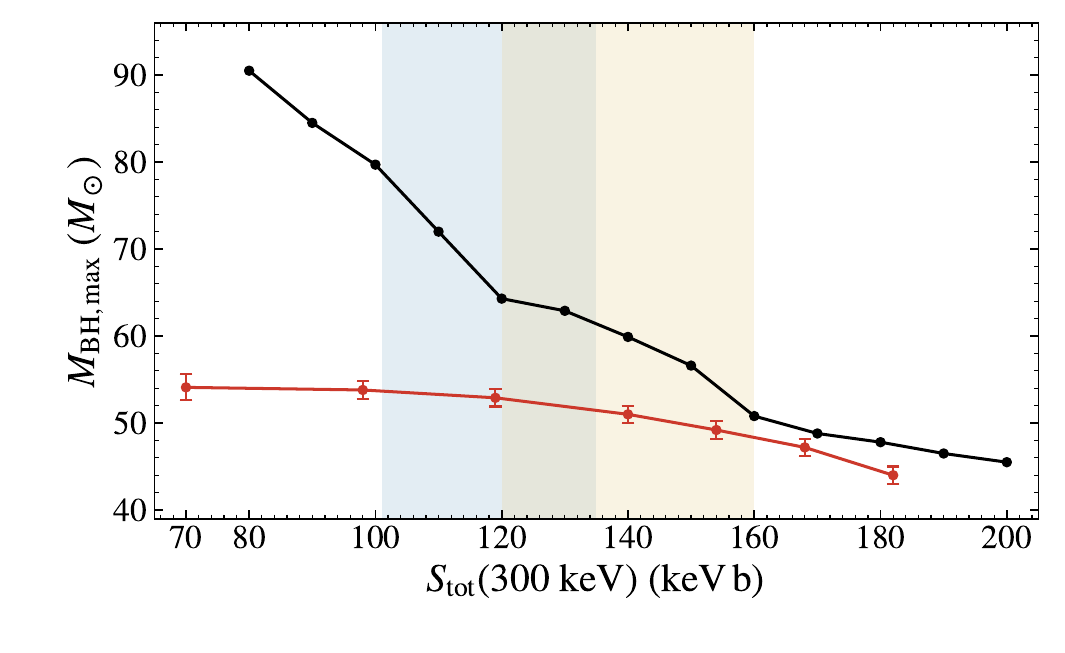}
\caption{Maximum black-hole mass below the pair-instability
mass gap as a function of $S_{\rm tot}(300\,\mathrm{keV})$.
The black curve shows the results of
Farag et al.~\cite{Farag2022} at $Z=10^{-5}$.
The red curve shows the largest selected bound stellar mass
at each reaction-rate factor from the present MESA
calculations at $Z=0.00142$.
These masses are provisional upper estimates of the final
black-hole masses because subsequent collapse was not followed.
For every red point of mass $M$, the upper error bar retains
the original assigned uncertainty, $\delta_{+}=\delta_1$, where
$\delta_1=1.5\,M_\odot$ for the leftmost point and
$1.0\,M_\odot$ for the others.
The lower error bar is
$\delta_{-}=\sqrt{\delta_{1}^2+(0.02M)^2}$,
combining the original uncertainty with an assumed downward
allowance of $2\%$ for possible mass loss during collapse.
The two contributions are treated as independent.
The central values of $M$ remain unchanged.
Thus, the error bars extend from $M-\delta_{-}$ to $M+\delta_{+}$.
The $2\%$ allowance was not calculated from the present models.
The blue band marks the adopted $101$--$135\,\mathrm{keV\,b}$
interval; the pale-yellow band marks the comparison interval
$120$--$160\,\mathrm{keV\,b}$.}
\label{fig:bhm_metallicity_compare}
\end{figure}

Figure~\ref{fig:bhm_metallicity_compare} compares the present
mass estimates at $Z=0.00142$ (red circles and line) with
the published results of Farag et al.~\cite{Farag2022}
at $Z=10^{-5}$ (black circles and line).
The Farag reaction-rate parameter is converted to
$S_{\rm tot}(300\,\mathrm{keV})$ using
Eq.~(\ref{eq:Farag_conversion}). The published points used
to interpolate within the adopted nuclear interval are
listed in Table~\ref{tab:Farag_points}.

The red curve gives smaller mass estimates than the black
curve throughout their common range of
$S_{\rm tot}(300\,\mathrm{keV})$.
This difference is consistent with stronger stellar winds
and greater mass loss at higher metallicity.
However, this comparison does not show how much of the mass
difference is caused by metallicity. The two studies also
use different initial stellar models, numerical settings,
and prescriptions for stellar mass loss. To isolate the
effect of metallicity, calculations would need to vary
metallicity while keeping these other assumptions fixed. 

\section{Connection between the two carbon-reaction problems}
\label{sec:connection}

The two reactions connect successive stages of stellar burning.
During helium burning, ${}^{12}\mathrm{C}(\alpha,\gamma){}^{16}\mathrm{O}$
helps determine how much carbon remains relative to oxygen.
Carbon fusion subsequently consumes carbon, releases energy,
and produces heavier nuclei.

The present carbon-fusion calculations examine two environments.
In an accreting white dwarf, the adopted rate changes the
conditions at which nuclear heating equals combined neutrino
and surface photon losses. Heating and neutrino cooling at
the center are also compared separately. In an initially
$20\,M_\odot$ star, the total fusion rate and its division
between the proton and alpha channels affect the abundance
profiles and isotope masses remaining at central carbon depletion.
The Yang proton-to-alpha ratio is applied only to this
massive-star comparison.

For helium capture on carbon, the composition left after
helium burning influences subsequent burning and pair-instability
pulses, and hence the mass available to form a black hole.
This dependence is examined using the published models at
$Z=10^{-5}$ and new MESA calculations at $Z=0.00142$.
These are separate stellar sequences; the two carbon-related
reaction rates are not varied together in a single calculation.

\section{Summary and conclusions}
\label{sec:summary}

The astrophysical consequences of
${}^{12}\mathrm{C}+{}^{12}\mathrm{C}$ fusion and
${}^{12}\mathrm{C}(\alpha,\gamma){}^{16}\mathrm{O}$
capture have been examined through their effects on stellar
heating, chemical composition, and black-hole masses.

For ${}^{12}\mathrm{C}+{}^{12}\mathrm{C}$, the Bayesian
thermonuclear rate is lower than the standard CF88 rate
over the temperature interval relevant to the white-dwarf
calculations. At fixed temperature, density, composition,
screening, and deposited energy per fusion event, carbon-fusion
heating is proportional to the reaction rate. In the MESA
sequences with $\dot M=10^{-7}\,M_\odot\,\mathrm{yr}^{-1}$,
the Bayesian rates move the condition
$L_{\rm nuc}=L_\nu+L_\gamma$ to larger white-dwarf masses
and higher central temperatures and densities.
For the Bayesian median, the central temperature and density
at this point are approximately $2.57\%$ and $18.9\%$ higher,
respectively, than in the calculation using the CF88 rate.
The differences among Bayes16, Bayes50, and Bayes84 are smaller
than their differences from the CF88 calculation.
The accretion-rate comparison shows that the balance density
is substantially more sensitive to $\dot M$ than the balance
temperature. A separate comparison identifies when nuclear
heating per unit mass at the center of the white dwarf equals
neutrino cooling per unit mass there,
$\epsilon_{\rm nuc,c}=\epsilon_{\nu,c}$.
This condition does not include heat transport or heating
caused by compression. Neither equality alone establishes
the onset of thermonuclear runaway. The Yang branching
ratio was not applied to the white-dwarf calculations.

For carbon burning in a star with an initial mass of
$20\,M_\odot$ and metallicity $Z=0.0142$, the calculation
using the CF88 fusion rate with MESA's standard division
between proton and alpha channels is compared with the
Bayesian-median rate using the Yang ratio.
At central carbon depletion, the largest relative changes
among the selected abundant products are $+10.78\%$ for
${}^{20}\mathrm{Ne}$ and $-28.70\%$ for ${}^{23}\mathrm{Na}$.
These differences include the combined effects of changing
the total rate and branching; their individual contributions
have not been separated. MESA's standard proportions for
the proton and alpha channels differ from those in the
original CF88 publication. The sensitivity of the results
to numerical resolution and to the slightly different
final central carbon abundances remains to be quantified.

For ${}^{12}\mathrm{C}(\alpha,\gamma){}^{16}\mathrm{O}$,
the nuclear input is taken from the $R$-matrix analysis
of Ref.~\cite{MukBHM2026}, which uses asymptotic normalization
coefficients to constrain bound-state contributions.
The narrow interval associated with these constraints,
$116$--$122\,\mathrm{keV\,b}$, does not include all systematic
uncertainties of the complete extrapolation.
The conservative interval
\begin{equation}
S_{\rm tot}(300\,\mathrm{keV})=118\pm17\,\mathrm{keV\,b}
\end{equation}
is therefore adopted. Applying it to the published
$Z=10^{-5}$ models of Farag et al.~\cite{Farag2022}
and interpolating between their tabulated results gives
\begin{equation}
61.5\,M_\odot\lesssim M_{\rm BH,max}\lesssim77.4\,M_\odot
\qquad (Z=10^{-5}).
\end{equation}
This range describes the maximum black-hole mass below
the pair-instability mass gap predicted by the published
$Z=10^{-5}$ stellar models of Farag et al., using their
adopted stellar-evolution assumptions. It reflects the
adopted nuclear interval and does not include all
stellar-model uncertainties.

The new MESA calculations at $Z=0.00142$ give smaller mass
estimates over the common range of
$S_{\rm tot}(300\,\mathrm{keV})$.
Most calculations were completed through the pulsational
mass-loss phase, yielding the bound stellar mass before
collapse. The subsequent collapse into a black hole was
not followed, so the plotted masses remain provisional
upper estimates of the final black-hole masses.
An assumed downward allowance of $2\%$ is combined in
quadrature with the original lower uncertainty at every
plotted point. The central values and upper error bars
remain unchanged. This allowance was not calculated
from the present models.

The smaller mass estimates at higher metallicity are
consistent with greater mass loss through stellar winds.
However, the two studies also use different initial stellar
models, numerical settings, and mass-loss prescriptions.
Calculations that vary metallicity while keeping these
other assumptions fixed are needed to determine its effect alone.

\appendix
\section{Numerical settings and endpoint properties}
\label{app:carbon_numerics}

The spatial mesh coefficient was set to \(1.0\), and the adaptive time-step control target was set to \(10^{-4}\) in both sequences. Calculations were stopped when the central carbon mass fraction fell below \(10^{-3}\). The revised sequence included additional stopping conditions at \(T_9=1.2\) for either the central temperature or the maximum temperature anywhere in the star. Carbon depletion was reached before either temperature limit was reached.

At the saved endpoints, the central carbon mass fractions were \(9.8657\times10^{-4}\) for the calculation using the CF88 rate and MESA branching and \(8.5030\times10^{-4}\) for the Bayesian-median sequence with Yang et al. branching. The corresponding central temperatures were \(9.6482\times10^8\) K and \(1.0150\times10^9\) K, and the final stellar masses were \(14.769844\,M_\odot\) and \(14.771465\,M_\odot\), respectively.

\section{Numerical implementation of the Yang et al. channel ratio}
\label{app:yang_implementation}

The proton-to-alpha thermonuclear rate ratio,
\begin{equation}
q(T_9)=\frac{R_p(T_9)}{R_\alpha(T_9)},
\qquad R_j=N_A\langle\sigma v\rangle_j,
\end{equation}
was obtained by digitizing the red curve in Fig.~8 of
Yang et al.~\cite{Yang}. The resulting numerical representation
is approximate and is not an author-supplied rate table.
Representative digitized values are $q\simeq0.293$, $0.452$,
and $0.527$ at $T_9=0.5$, $1.0$, and $1.2$, respectively.
These values are quoted to identify the adopted curve, rather
than to indicate its numerical accuracy.

For adjacent digitized points $(T_{9,k},q_k)$ and
$(T_{9,k+1},q_{k+1})$, the ratio was evaluated by linear
interpolation in temperature,
\begin{equation}
q(T_9)=q_k+
\frac{T_9-T_{9,k}}{T_{9,k+1}-T_{9,k}}
(q_{k+1}-q_k).
\label{eq:yang_linear_interpolation}
\end{equation}
This prescription applies within the tabulated range.
Below $T_9=0.5$, the ratio was held fixed at $q=0.29$;
this continuation is an explicit low-temperature extrapolation.

Let $R_{\alpha+p}^{(50)}(T_9)$ denote the sum of the alpha
and proton rates in the original Bayesian-median input.
At each tabulated temperature, the revised channel rates were
constructed as
\begin{align}
R_\alpha^{(50,Y)}(T_9)
&=\frac{R_{\alpha+p}^{(50)}(T_9)}{1+q(T_9)},\\
R_p^{(50,Y)}(T_9)
&=\frac{q(T_9)R_{\alpha+p}^{(50)}(T_9)}{1+q(T_9)}.
\label{eq:yang_appendix_repartition}
\end{align}
The construction preserves the alpha-plus-proton sum and
reproduces the adopted ratio at each table entry.
No neutron channel was included in the adopted Bayesian
carbon-fusion input.
The reverse alpha and proton rates were multiplied by
the same respective factors as their forward rates.
The Yang et al. ratio was applied only to the revised sequence;
the calculation using the CF88 rate and MESA branching retained its original channel branching.

Yang et al. evaluated their thermonuclear channel ratio using
an energy-independent total-fusion modified astrophysical
factor, $S_{\rm tot}^{*}=3\times10^{16}\,\mathrm{MeV\,b}$,
following CF88~\cite{Yang,CF88}. Applying their ratio to the
Bayesian-median total assumes that the thermally averaged
ratio remains applicable with the different energy dependence
of the Bayesian nuclear input. The individual channel rates
were not recalculated by integrating the energy-dependent
channel fractions with that input. Uncertainties from the
digitization and this transfer of the ratio were not quantified;
the procedure does not yield separate Bayesian probability
distributions for the channel rates.


\begin{thebibliography}{99}

\bibitem{Bennett2012}
M.~E.~Bennett et al.,
The effect of $^{12}\mathrm{C}+{}^{12}\mathrm{C}$ rate uncertainties
on the evolution and nucleosynthesis of massive stars,
Mon. Not. R. Astron. Soc. \textbf{420}, 3047 (2012),
doi:10.1111/j.1365-2966.2012.20193.x.

\bibitem{Pignatari2013}
M.~Pignatari et al.,
The $^{12}\mathrm{C}+{}^{12}\mathrm{C}$ reaction and the impact
on nucleosynthesis in massive stars,
Astrophys. J. \textbf{762}, 31 (2013),
doi:10.1088/0004-637X/762/1/31.

\bibitem{Chieffi2021}
A.~Chieffi et al.,
Impact of the New Measurement of the
$^{12}\mathrm{C}+{}^{12}\mathrm{C}$ Fusion Cross Section on
the Final Compactness of Massive Stars,
Astrophys. J. \textbf{916}, 79 (2021),
doi:10.3847/1538-4357/ac06ca.

\bibitem{Monpribat2022}
E.~Monpribat et al.,
A new $^{12}\mathrm{C}+{}^{12}\mathrm{C}$ nuclear reaction rate:
Impact on stellar evolution,
Astron. Astrophys. \textbf{660}, A47 (2022),
doi:10.1051/0004-6361/202141858.

\bibitem{Gesue}
R.~M.~Gesu\`e, O.~Straniero, G.~Imbriani, A.~Di Leva, A.~Best,
R.~Buompane, K.~Chakraborty, I.~Dom\'inguez, F.~Ercolano,
L.~Gialanella, L.~Morales-Gallegos, D.~Rapagnani, M.~Romoli,
S.~Khan, C.~Santonastaso, and R.~P.~Yadav,
Triggering carbon burning in stars: Main reaction rates revised?,
Astron. Astrophys. \textbf{714}, A448 (2026),
doi:10.1051/0004-6361/202660163.

\bibitem{Chieffi2025}
A.~Chieffi et al.,
The status of $^{12}\mathrm{C}+{}^{12}\mathrm{C}$ fusion process,
Eur. Phys. J. A \textbf{61}, 280 (2025),
doi:10.1140/epja/s10050-025-01759-3.

\bibitem{NanWang2025}
W.~Nan, Y.~Wang, J.~Su et al.,
Determination of resonances in Gamow window for
$^{12}\mathrm{C}+{}^{12}\mathrm{C}$ fusion reaction via
thick-target inverse kinematics method,
Phys. Lett. B \textbf{862}, 139341 (2025),
doi:10.1016/j.physletb.2025.139341.

\bibitem{MukCC2026}
A.~M.~Mukhamedzhanov,
Bayesian smooth-fit extrapolation of the
$^{12}\mathrm{C}+{}^{12}\mathrm{C}$ astrophysical $S$ factor,
Phys. Rev. C \textbf{114}, 015809 (2026), doi:10.1103/vhms-qgr4.

\bibitem{Yang}
R.~Yang, R.~Chen, X.~Fang, Y.~Fan, X.~Tang, Y.~Li, and F.~Luo,
Proton-to-alpha branching ratio in the
$^{12}\mathrm{C}+{}^{12}\mathrm{C}$ fusion reaction at astrophysical energies,
Chin. Phys. C \textbf{50}, 094003 (2026),
doi:10.1088/1674-1137/ae823a.

\bibitem{CF88}
G.~R.~Caughlan and W.~A.~Fowler,
Thermonuclear reaction rates V,
At. Data Nucl. Data Tables \textbf{40}, 283 (1988).

\bibitem{GesueData}
R.~M.~Gesu\`e et al.,
Triggering carbon burning in stars. Main reaction rates revised,
Zenodo dataset (2026), doi:10.5281/zenodo.22875649.

\bibitem{deBoer2017}
R.~J. deBoer, J. G\"orres, M. Wiescher, R.~E. Azuma, A. Best,
C.~R. Brune, C.~E. Fields, S. Jones, M. Pignatari, D. Sayre,
K. Smith, F.~X. Timmes, and E. Uberseder,
Rev. Mod. Phys. \textbf{89}, 035007 (2017).

\bibitem{Farmer2020}
R.~Farmer, M.~Renzo, S.~E.~de Mink, M.~Fishbach, and S.~Justham,
Constraints from gravitational-wave detections of binary black hole
mergers on the $^{12}\mathrm{C}(\alpha,\gamma)^{16}\mathrm{O}$ rate,
Astrophys. J. Lett. \textbf{902}, L36 (2020).

\bibitem{WoosleyHeger2021}
S.~E.~Woosley and A.~Heger,
The Pair-instability Mass Gap for Black Holes,
Astrophys. J. Lett. \textbf{912}, L31 (2021),
doi:10.3847/2041-8213/abf2c4.

\bibitem{Farag2022}
E.~Farag, M.~Renzo, R.~Farmer, M.~T.~Chidester, and F.~X.~Timmes,
Resolving the peak of the black hole mass spectrum,
Astrophys. J. \textbf{937}, 112 (2022),
doi:10.3847/1538-4357/ac8b83.

\bibitem{Wang2026}
Y.-Z.~Wang, Y.-J.~Li, S.-J.~Gao, S.-P.~Tang, and Y.-Z.~Fan,
A new group of low-spin $50M_\odot$--$70M_\odot$ black holes
and the high pair-instability mass cutoff,
Sci. China Phys. Mech. Astron. \textbf{69}, 299562 (2026),
doi:10.1007/s11433-026-3004-6.

\bibitem{MukBHM2026}
A.~M.~Mukhamedzhanov,
Nuclear constraints on $^{12}\mathrm{C}(\alpha,\gamma)^{16}\mathrm{O}$
and their impact on black-hole mass predictions,
Phys. Rev. C \textbf{114}, 035803 (2026),
doi:10.1103/hy6l-43l3.

\bibitem{Mehta2022}
A.~K.~Mehta, A.~Buonanno, J.~Gair, M.~C.~Miller, E.~Farag,
R.~J.~deBoer, M.~Wiescher, and F.~X.~Timmes,
Observing intermediate-mass black holes and the upper stellar-mass gap
with LIGO and Virgo,
Astrophys. J. \textbf{924}, 39 (2022).

\bibitem{Woosley2017}
S.~E.~Woosley,
Pulsational Pair-instability Supernovae,
Astrophys. J. \textbf{836}, 244 (2017),
doi:10.3847/1538-4357/836/2/244.

\bibitem{Woosley2002}
S.~E.~Woosley, A.~Heger, and T.~A.~Weaver,
The evolution and explosion of massive stars,
Rev. Mod. Phys. \textbf{74}, 1015 (2002),
doi:10.1103/RevModPhys.74.1015.

\bibitem{Paxton2011}
B.~Paxton, L.~Bildsten, A.~Dotter, F.~Herwig, P.~Lesaffre,
and F.~Timmes,
Modules for Experiments in Stellar Astrophysics (MESA),
Astrophys. J. Suppl. Ser. \textbf{192}, 3 (2011).

\bibitem{Jermyn2023}
A.~S.~Jermyn et al.,
Modules for Experiments in Stellar Astrophysics (MESA):
Time-dependent convection, energy conservation, automatic
 differentiation, and infrastructure,
Astrophys. J. Suppl. Ser. \textbf{265}, 15 (2023).

 \bibitem{Rahman2022}
N.~Rahman, H.-T.~Janka, G.~Stockinger, and S.~Woosley,
``Pulsational pair-instability supernovae: gravitational
collapse, black hole formation, and beyond,''
Mon. Not. R. Astron. Soc. \textbf{512}, 4503--4540 (2022),
doi:10.1093/mnras/stac758.

\end{thebibliography}
\end{document}